# Teaching with a companion: the case of gravity

Iuliia Zhurakovskaia
University Paris-Saclay
CNRS, LISN
Orsay, France
iuliia.zhurakovskaia@gmail.com

Jeanne Vezien
University Paris-Saclay
CNRS, LISN
Orsay, France
jeanne.vezien@limsi.fr

Patrick Bourdot
University Paris-Saclay
CNRS, LISN
Orsay, France
patrick.bourdot@limsi.fr

*Abstract*— Virtual Reality (VR) has repeatedly proven its effectiveness in student learning. However, despite its benefits, the student equipped with a personal headset remains isolated from the real world while immersed in a virtual space and the classic student-teacher model of learning is difficult to transpose in such a situation. This study aims to bring the teacher back into the learning process when students use a VR headset. We describe the benefits of using a companion for educational purposes, taking as a test case the concept of gravity. We present an experimental setup designed to compare three different teaching contexts: with a physically present real teacher, using a live video of the teacher, and with a VR avatar of the teacher. We designed and evaluated three scenarios to teach the concept of gravity: an introduction to the concept of free fall, a parabolic trajectory workshop and a final exercise combining both approaches. Due to sanitary conditions, only pre-tests are reported. The results showed that the effectiveness of using the VR simulations for learning and the self-confidence level of the students increased as well. The interviews show that the students ranked the teaching modes in this order: VR companion mode, video communication and real teacher.

*Keywords-virtual reality; companion; science education*

I. INTRODUCTION: COMPANIONS IN THE DIGITAL ERA

Humans are social animals and usually prefer to avoid being alone whenever possible. During our studies we are usually surrounded by other people, students, and professors.

On the other hand, VR has been democratised recently on the advent of affordable, personal devices - head-mounted displays (HMDs). These systems are very immersive, but they considerably reduce classical social interactions, which poses a problem when using them for an educational purpose. For example, students often need to ask questions during their educational process. Questions arise due to a lack of attention, difficulties in understanding certain concepts, or even misconceptions on the subject. While wearing an HMD, it can be complicated to switch from VR to the real world to get assistance, then to go back to work in VR. The availability of virtual help - a companion, can solve the problem [2]. This is the rationale for the present work. Thus, our research question is "Given the isolation problem of VR headsets, how can we best bring back the teacher in the learning process?".

Companions are an evolution of Embodied Conversational Agents (ECA). They are endowed with emotional abilities that make them capable of establishing social and affective relationships with people, caring for them and providing them with companionship. The latter is recognised as a key factor in children's acquisition and development of social skills [3].

The use of pedagogical agents in 3D simulations can be traced to Rickel and Johnson [22] creating the STEVE (Soar Training Expert for Virtual Environments) agent. STEVE was used as a teaching assistant during procedural learning, providing verbal or direct cues to students upon requests.

This type of pedagogical agent, represented by an animated character, is now described as ABLEs (Agent-Based Learning Environments) [23]. Research has found that the visual presence of such personal assistance does not distract students from learning [24]. The presence of a digital companion has been found to have a positive effect on learners' perception of their learning experience as well as their performance outcomes with these environments [25].

These works show that companions can be quite effective, but their appearance and behaviour can affect this effectiveness in the teaching process. Other works do not discuss how the companion is rendered, and instead focus on the AI (Artificial Intelligence) aspect. Since these companions are designed to replace the physical teacher in a VR-based serious game, it made sense to investigate how to include them in existing gaming education framework. Our research is complementary to these "automated" approaches.

Companions have been used in a few studies related to education. Firstly, it has been suggested by Lea Pillette et al. [4] that users may experience difficulties or be nervous during training sessions using Mental-Imagery Brain-Computer Interfaces (MI-BCI) and that this may be partly the result of a lack of social presence and emotional support, which had been paid little attention to. One way of providing this social and emotional context is using a learning companion. PEANUT (Personalised Emotional Agent for Neurotechnology User Training) provided social presence and emotional support depending on the performance and progress of the user through interventions combining both spoken sentences and facial expressions. Pillette et al. [4] found desirable characteristics of their learning companion in terms of verbal content and appearance (e.g., eyebrows can increase the expressiveness of cartoon faces). They found that *«non-autonomous people (those who are more inclined to work in a group), who are usually at a disadvantage when using MI-BCI, benefited compared to autonomous people with PEANUT with a 3.9% increase in peak performance. In addition, in terms of user experience, PEANUT seems to have improved by 7.4% people's attitude towards their ability to learn and remember how to use MI-BCI, which is one aspect of the user experience we evaluated»*.

N. Roa-Seïler conducted a study regarding the presentation of companions [3]. In this study, they compared which companion the 4th-grade children preferred in maths class: Samuela is a 3D screen-based character, Nao is an actual robot, and the third embodiment was Ari, an actual person made to look like a cartoon character and projected on the screen. The study used a Wizard of Oz setup (WoZ), where the teacher controlled the companions using a panel with predefined sets of action. The children marked Ari as the most affectionate because of her behaviour ("She smiled", "She talked nice"). Nao, not having a mouth, couldn't smile, which generated some confusion; nevertheless, children found him fascinating probably because none of them had met a robot before, whereas they had already encountered screen-based characters in video games.

As a support of our work on companions, we decided to focus on the topic of gravity, as it is one of the subjects in physics that students often have misconceptions about.

## II. TEACHING GRAVITY WITH COMPANIONS

### A. Why free-fall resist teaching

What do an apple and a planet have in common? They are both subject to the same force that describes their movement: gravity. Gravity is the fundamental force that we experience first, as we see it in action everywhere.

In the French education system, the official teaching of free-fall is based on a progression that consists of establishing the Cartesian equation of a projectile trajectory, deducing the motion's nature, and then demonstrating why a projectile moving in a homogeneous field of gravity is a conservative system. But this approach to parabolic motion is insufficient for a complete and consistent understanding of free-fall, especially regarding students' representations of projectile trajectories in a gravitational field, launched under different initial conditions (initial velocity vector) [5]. This suspicion is confirmed by A. Prescott [6], who showed that the parabola is not the preferred curve for students to represent free-fall motion (even after learning).

Halloun and Hestenes [7] note that "*students hold the pre-scientific belief that every movement has a cause*". Also, A. Prescott [6] showed that students have "*misconceptions when faced with a projectile movement situation*". Most students do not consider the initial velocity and draw a straight line when the trajectory of motion crosses the equilibrium position (the result is not a parabola but two upper lines of a triangle) [8]. Thus, it is clearly necessary to dispel students' misconceptions on free-fall while teaching them parabolic motion.

Where do these errors come from? It is generally accepted that we think the way we have been taught to think. Nevertheless, regarding physics phenomena, we all start with "intuitive physics": a kind of general explanatory scheme representing a common and self-consistent set of concepts and resisting attempts to change or modify it, no matter how wrong it may be. This is not taught but comes unconsciously from our already existing knowledge and experience. These intuitive patterns are a strong obstacle to actual scientific teaching. It is precisely this "intuitive physics" that prevents us from gaining new knowledge and makes teaching less effective [9].

Spontaneous reasoning is very stable and not susceptible to learning that contradicts it. Just like delusions, they must first be destroyed before being replaced [10]. It is not just "a few mistakes" by students, it is a way of thinking found even in everyday conversations [9]. Thus, one must first show students the limitations of their current reasoning and only then, teach new material. Learning becomes a process in which new concepts have to displace or rearrange stable concepts that the students have been building up for a long time.

When studying free-fall, the most common misunderstanding is the relationship between force and acceleration (force and motion) [11] and between position, velocity and acceleration in one dimension [12]. These Newtonian concepts are often distorted or misinterpreted by students to fit their ingrained misconceptions. Students tend to memorise them separately, as formulas, with little or no connection to fundamental qualitative concepts. Put simply, students cannot apply their theoretical knowledge to practical tasks. This is aggravated by the fact that the modern education system often puts an emphasis on quantitative calculations to measure learning outcomes, instead of assessing true qualitative understanding.

### B. Teaching gravity

L. McDermott [12] found that results obtained with students at the university level, including prospective and practising teachers, are equally applicable to high school students and, in some cases, to primary school children. Thus, we can conclude that students are pulling their delusions regarding physics from their earliest years. In a survey of 6000 high school, college and university students who took the Force Concept Inventory test (designed to assess students' understanding of Newtonian dynamics [13] [14]) before and after learning mechanics, it was found that the largest gains in scores occurred in those students who were involved in interactive activities that provided immediate feedback through discussion with peers or instructors [15] [12].

This raises the question of how to teach students more effectively. When students have misunderstandings about the differences between instantaneous velocity and constant acceleration, a common teaching strategy to help students overcome some conceptual difficulties is to use Microcomputer-Based Laboratory work (MBL) [12] [18]. In kinematics, students plot in real time the relationship between position, velocity and acceleration versus time for motions, including their own. Instantaneous feedback helps to explain the links between movements and their graphical representations. An introductory course should be based entirely on such laboratory work [12]. Evaluation of the syllabus by means of pre- and post-tests shows that achievement and memorisation are then significantly higher than in courses taught by traditional methods. In another type of lab approach, students perform simple experiments that are designed to form the basis of Socratic dialogue [16] [17].

Thus, questions that require qualitative reasoning and verbal explanation are very important. Mathematical formalism should be set aside until students have had some practice in qualitative reasoning about the phenomena being studied. Moreover, students should be asked to synthesise concepts and mathematical deductions and to formulate relationships in their own words. Likewise, learners should engage in the process of constructing qualitative models that will help them understand the connections and differences between concepts. Persistent conceptual difficulties should be explicitly overcome by repeated challenges in different contexts. Based on these findings, teaching by telling is an ineffective teaching method for most learners. Students need to be intellectually active in order to develop functional understanding. In conclusion, while teaching the concept of free fall in a gravity field, interaction on the one hand, and dialogue (questions and answers) on the other hand, seem to play a key role in the learning mechanism and in dispelling conceptual errors.

## III. EXPERIMENTS

### A. Research questions

VR naturally responds to the interaction aspects by creating arbitrary experimental conditions. On the other hand, when a student is in an immersive situation via a HMD, they are cut off from the outside world, particularly from the teacher, which tends to diminish or even interrupt the crucial questioning and challenging mechanisms. As the process of learning without a teacher can be difficult, this study's main interrogation concerns the isolation problem created using VR systems in an educative framework. Several objectives can be formulated: (1) The need to bring the teacher back into the virtual educational process and how best to do so; (2) help students understand and overcome their mistakes and misconceptions while learning free-fall; (3) convey to students the key idea of parabolic motion and how it can change in the presence or absence of gravitational fields.

*B. Experimental conditions*

We tested three playing conditions in a VR setting:

**Condition 1:** The student is alone in the virtual world. The participant is equipped with a VR headset, two 3D controllers and a set of headphones. To talk to the teacher, they must remove the headset and address the teacher in person.

**Condition 2:** Student plays in VR but stays in contact with the teacher via an audio-visual link (videoconferencing window inside the simulation) (Fig.1). This way the student can see and hear the teacher in real time, without breaking immersion.

**Condition 3:** The student experiences the VR together with a companion, a robot named Kylie (Fig.3). Kylie will follow him everywhere. The teacher can manage this companion as WoZ, using a set of predefined actions (for example, making the robot point to the object that the student should pay attention to). The teacher can address the student with a microphone, and the student perceives the voice as coming from the robot. We chose a robot model because it is human-like [19], and more relatable than, say, a bird or a dragon. We chose to make it gender-neutral and did not try to replicate the teacher's appearance, because a teacher maybe be unwilling or unable to create a resembling avatar in an actual classroom condition.

*C. Interviews with teachers*

Prior to conducting the experiments, three interviews were conducted with teachers: two primary school teachers and a middle school English teacher. Each of them had more than 30 years of work experience. During the interview they got descriptions of the three game conditions and asked to choose what they would prefer and why. They were also asked how they generally felt about this method of teaching with VR.

Two teachers (one primary school teacher and an English teacher) preferred the second condition. They justified this by saying that in this case the pupil does the whole experience by himself/herself and if he/she needs help, face-to-face communication will be more effective for getting all the information. They also pointed out that face-to-face communication plays a big role in teaching and pupils should not be left to their own devices. Indeed, children in primary school may have comprehension problems due to inattention, and they are more likely to listen to the teacher (and take his face as authority) than to follow the robot's instructions. To the second question, the primary school teacher answered that in her practice this method of teaching would not be useful, again because the children were too young and did not have the necessary autonomy. The English teacher answered positively that she would use such a method in her teaching practice (close to distance learning, which the Covid-19 crisis has made commonplace). Another primary school teacher chose the first condition. She pointed out that it is necessary to combine modern technology and face-to-face communication and not to detach from social communication in general. She added that she would not use such technology in mainstream classes, but perhaps in supplementary courses. She suggested that this method could be applied to older students (middle and high school). In addition, she and her children are already using 2D simulations on tablets. According to her, today's children love technology, but human interaction must remain at the forefront.

From these interviews, we can hypothesise that there will be a strong preference towards "video" or "real teacher" conditions during actual experiments. But it also points out that, should the VR technology be spread in classrooms, teachers could be reluctant to use it for fear of losing crucial social interactions with their pupils.

*D. Experimental setup*

As mentioned, this study involves three exercises. The exercises are proposed one after the other in this order:

*1) Exercise 1 - Gravitation*

The goal of this exercise is to show how gravity makes objects fall. This exercise is specifically designed to address the most common misconception in this regard, i.e., that heavy objects fall faster than light ones.

In this exercise, the student has 9 objects of different weights, sizes and densities at their disposal, as well as a Roberval balance scale [1]. The experience goes as follows: the student grabs any two objects from the table, compares their weights on the scale, places these two objects on the two predefined target locations and presses the "Start" button. The objects fall down, and a timer displays the time they each take to fall to the floor (Fig.1).

Three game locations are proposed in succession: Earth ($g=9.807 m/s^2$), Moon ($g=1.62 m/s^2$), outer space (no gravity). The visuals of the background change accordingly.

*2) Exercise 2 - Parabolic trajectory*

The purpose of this exercise is to show that in the presence of gravity, the fly trajectory of the projectile will always be parabolic in a gravity field - strongly bent, bell-shaped under Earth's gravity, less so on the Moon, and a straight line in the absence of gravity, contrary to what many students will spontaneously express.

In this game the student can choose between two types of projectiles (1 kg and 2 kg) to feed a cannon and shoot targets. When the cannon is loaded, the trajectory corresponding to the current cannon orientation is displayed in the form of a string of small beads. The student can observe the flight path in 3D live view and on a 2D orthographic monitor (Fig.2).

The student can take up to 8 shots in each gravity condition. For each attempt, the target is randomly placed in one out of 8 positions at different distances and elevations.

*3) Exercise 3 - Parabolic trajectory & Gravity*

In this exercise, the student can observe how the trajectory of the projectile changes depending on the presence or absence of a gravitational field. Two kinds of cubic areas, "G - gravity" (normal earth gravity) and "NG - no gravity", called *boxes*, are horizontally stacked in a sandwich manner: on one platform "G-NG-G" and on another "NG-G-NG" (Fig.3).

Again, the main task consists in shooting a cannonball. The student is asked to focus on the shape of the trajectory as the ball traverses the "gravity sandwich" (no targets). Before each shot, the student sees only the beginning of the predicted trajectory before the gravity sandwich. There are eight 1 kg projectiles on each platform. The student can watch the full trajectory when the shot is fired on a 2D orthographic video monitor.

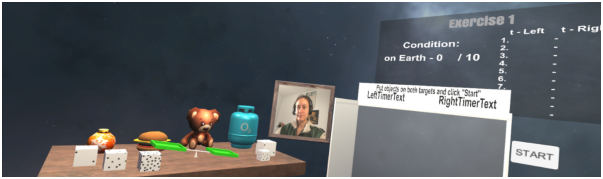

Figure 1. Exercise 1 – Gravitation (Condition 2).

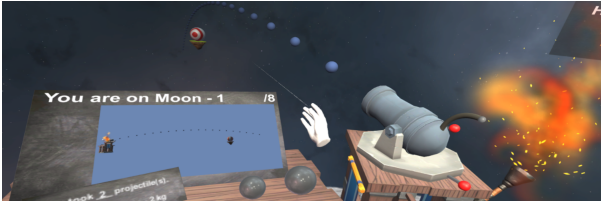

Figure 2. Exercise 2 – Parabolic trajectory (Condition 1).

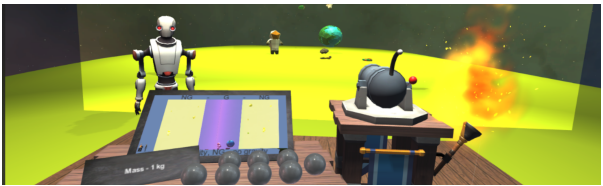

Figure 3. Exercise 3 – Parabolic trajectory and Gravity (Condition 3).

*E. Questionnaire*

In order to evaluate the students' comprehension of the subject of gravity and free-falling, we developed our own questionnaire consisting of nine questions accompanied by a self-assessment of the level of confidence in answering the question. This questionnaire focuses on misconceptions and qualitative knowledge (link to the questionnaire: https://figshare.com/s/0902f3b748f76b3fb2ed).

*F. Experimental procedure*

Proper sanitary procedures to prevent Covid-19 contamination were designed and applied [20] during the experiments, described below.

Users passed a pre-test questionnaire. Players received basic practical instructions on how to interact with objects in VR. Then the player starts with Exercise 1. In each mode and for each exercise, instructions regarding the course of actions were given by the teacher by voice, but the general task description was also duplicated on a virtual whiteboard inside the simulation.

Following the VR experiment, the participants passed a post-test (the same as the pre-test). Then, participants were then asked to undergo a 15-minutes semi-directed interview, with the objective to collect some data regarding the subjective perception of the overall quality of the environment and interaction, to justify some answers in the post-test (if misconceptions remained). We were especially interested in feedback regarding the presence of the helper. Finally, students were given a theoretical description of each of the three game modes and asked to choose which one they would ideally prefer, and why.

## IV. PRELIMINARY RESULTS

Due to the sanitary situation caused by the Covid-19 crisis, a complete round of experiments could not be undertaken at the time of writing this document. However, a preliminary study was conducted.

There were six participants (Mean age = 27, SD = 2; 2 females and 4 males; 2 per condition). All participants were right-handed, without visual problems, all have experience of using VR. All had a level equivalent to Masters in a science-related field. Four students said that they had last studied the concept of gravity in secondary school, two others in high school.

During the game we recorded the time of completion for each condition and for each exercise. Because the aim was to study knowledge acquisition, success rate (target hitting skills) was not recorded. We consequently focused on test results and personal interviews.

TABLE I. TIME SPENT IN THE GAME (IN MINUTES)

|  | Cond. 1 | Cond. 2 | Cond. 3 | Average t |
|---|---|---|---|---|
| Ex.1 | 9,33 | 14,07 | 15,45 | 12,95 |
| Ex.2 | 6,79 | 8,75 | 10,09 | 8,54 |
| Ex.3 | 3,62 | 4,53 | 6,92 | 5,03 |
| Average total | 19,74 | 27,36 | 32,45 | 26,52 |

Table 1 compiles the total time results for each mode, for each state, and the total playing time. The average playing time was 26.52 minutes. However, for the real time mode students spent 19.74 minutes, for the video mode it took longer at 27.36 minutes, and the longest time was obtained for the companion mode at 32.45 minutes. On average students spent the most time on the 1st exercise with 12.95 minutes, followed by the second exercise with 8.54 minutes and the shortest time was for the third exercise with 5.03 minutes. The first and second exercise had 3 states, so there was a familiarity factor: the first exercise took the longest time, the second was shorter, and the third the shortest (for 5 out of 6 participants). For Exercise 3, the same amount of time was spent on average for each platform (2.12 minutes and 2.01 minutes).

Based on the Pre- and Post-test questionnaire success rate, all participants improved their performance, with average success rising from 77.56% to 87.82%. The self-confidence score increased by 13.42% as well. Looking at success rates for each mode, companion mode was the most effective, with an increase of 17%, followed by no-help mode with an increase of 8%, and finally video mode with 6%. However, these results need to be confirmed with a more statistically significant set of participants.

For the most part, scores increased on questions that were directly in relation to the game. In contrast, answers to question related with determining the flight path of a ball if dropped from an airplane (following a parabolic trajectory, but not directly considered in the practical exercises), remained consistently wrong in most cases. It seems the notion of initial speed is not invariant depending on how it is acquired (cannonball vs. dropping from a moving object), so that parabolic "forward" motion is not generalised easily.

During the interview, all the students noted the simplicity of Exercise 1, but one participant did not believe at first that 2 objects with different masses could fall the same way and concluded that the program did not work correctly, then later changed his mind. In Exercise 2 half of the students commented on the usefulness of the 2D orthographic view, the other half found the exercise was obvious without it, but all enjoyed shooting at targets with a cannon. Exercise 3 was also evaluated positively, but half of the students regretted that it was not possible to shoot at specific objects. In this situation, the 2D orthographic side view was deemed useful because the transition between the gravity blocks was not visible from the participant's subjective view.

When asked to rank the exercises by order of preference, half of the students put Exercise 2 in first place and the other half chose Exercise 3, but all students clearly put Exercise 1 in last place, justifying this by the over simplicity.

When asked about their preferred assisting mode, five students chose the companion mode, justifying that it was more interesting, that they preferred not to see the teacher in the real world because they would feel pressure, and that interacting with the companion was more useful as it could show things in a 3D context without ambiguity. One student chose video communication, saying he preferred to see the teacher, i.e. that one-to-one communication was important. One student noted that the choice of game mode can depend on different factors: for instance, in a short simulation in VR the teacher was not required, whereas a long tutorial lesson would certainly make interacting with the teacher a necessity.

Students who tested the first mode (with a real teacher) noted that they had difficulty understanding what the teacher wanted from them. Being completely immersed in the VR world, an "outside" voice was hard to focus on. This effect was absent in the second and third mode subjects.

Obviously further experiments with a higher number of participants are necessary to confirm our findings.

## V. Conclusion

Many students have misconceptions about what could be considered the most basic concepts of physics. Although the use of VR itself for learning has proven to be effective [21], the fact remains that the complete teaching process cannot be realised in a virtual environment alone. Often the presence of a teacher is a key factor. Because being immersed in a virtual world tend to isolate the student from the real world, a virtual companion or live video link with the teacher can significantly counterbalance this situation. Our goal was thus to study the effectiveness of assisting techniques in serious VR games. The increased average success rate of all participants of the preliminary tests confirms that indeed such assistance complemented the natural benefits of VR learning.

Preliminary studies demonstrate the ecological validity of VR, giving access to situations that would otherwise never be experienced. Based on success rates for each mode, we found that companion mode was the most effective.

## Acknowledgement

This work was supported by French government funding under grant ANR-21-ESRE-0030 / CONTINUUM.